\documentclass[12pt,nohyper,letterpaper]{JHEP3}         
\usepackage{amssymb,amsfonts}

\usepackage{epsfig}

\def\be{\begin{eqnarray}}
\def\ee{\end{eqnarray}}
\newcommand{\nn}{\nonumber}
\newcommand\para{\paragraph{}}

\newcommand{\eqn}[1]{(\ref{#1})}

\def\Dslash{\,\,{\raise.15ex\hbox{/}\mkern-12mu D}}
\def\Dbarslash{\,\,{\raise.15ex\hbox{/}\mkern-12mu {\bar D}}}
\def\delslash{\,\,{\raise.15ex\hbox{/}\mkern-9mu \partial}}
\def\delbarslash{\,\,{\raise.15ex\hbox{/}\mkern-9mu {\bar\partial}}}
\def\pslash{\,\,{\raise.15ex\hbox{/}\mkern-9mu p}}
\def\calDslash{\,\,{\raise.15ex\hbox{/}\mkern-12mu {\cal D}}}

\newcommand{\ket}{\rangle}
\newcommand{\Tr}{{\rm Tr}}

\def\lae{\mathrel{\mathop{\smash{\lower .5 ex \hbox{$\stackrel<\sim$}}}}}
\def\lae{\mathrel{\mathop{\smash{\lower .5 ex \hbox{$\stackrel>\sim$}}}}}

\def\theequation{\thesection.\arabic{equation}}

\preprint{DAMTP-2007-84}

\title{The Geometric Phase in Supersymmetric Quantum Mechanics}
\author{Chris Pedder, Julian Sonner and David Tong\\
Department of Applied Mathematics and Theoretical Physics, \\
University of Cambridge, UK\\{\tt c.j.pedder, j.sonner,
d.tong@damtp.cam.ac.uk}}

\abstract{We explore the geometric phase in ${\cal N}=(2,2)$
supersymmetric quantum mechanics. The Witten index ensures the
existence of degenerate ground states, resulting in a non-Abelian
Berry connection. We exhibit a non-renormalization theorem which
prohibits the connection from receiving perturbative corrections.
However, we show that it does receive corrections from BPS
instantons. We compute the one-instanton contribution to the Berry
connection for the massive ${\bf CP}^1$ sigma-model as the
potential is varied. This system has two ground states and the
associated Berry connection is the smooth $SU(2)$ 't
Hooft-Polyakov monopole.}

\begin{document}
\pagestyle{plain} \setcounter{page}{1}
\newcounter{bean}
\baselineskip16pt

\section{Introduction}

The purpose of this paper is to explore the geometric phase --- or
Berry's phase ---  among the ground states of quantum mechanical
systems exhibiting ${\cal N}=(2,2)$ supersymmetry. Our goal is to
show that Berry's phase and supersymmetry are natural bedfellows:
the Berry connection is protected by a perturbative
non-renormalization theorem, but receives corrections from BPS
instantons.

\para
Berry's phase governs the evolution of a quantum state as the
parameters of the system are varied adiabatically
\cite{berry,simon,wz}. Consider a Hamiltonian $H(\vec{m})$
depending on the collection of parameters $\vec{m}\in{\cal M}$. We
focus on the fate of the $N$ ground states of the system, spanned
by the basis $|\psi_a(\vec{m})\rangle$, $a=1,\ldots,N$. As the
parameters vary adiabatically along a closed path $\Gamma$ in
${\cal M}$, the ground states return to themselves up to a $U(N)$
rotation,
\be |\,\psi_a\rangle \rightarrow P\,\exp\left(-i\oint_\Gamma
\vec{A}_{ab}\cdot d\vec{m}\right)\,|\,\psi_b\rangle\ee
where the $u(N)$ valued Berry connection over ${\cal M}$ is
defined by
\be \vec{A}_{ab} = i\langle \psi_b |\frac{\partial}{\partial
\vec{m}} |\,\psi_a\rangle \label{time}\ .\ee
The canonical example of an Abelian Berry's phase arises for a
spin 1/2 particle in a magnetic field $\vec{B}$
\be H = \vec{B}\cdot \vec{\sigma}\label{1st}\ee
where $\vec{\sigma}$ are the Pauli matrices and the magnetic field
$\vec{B}\in {\bf R}^3$ plays the role of the parameters $\vec{m}$.
It is a simple matter to compute the Abelian Berry connection
$\vec{A}$ for the ground state of this system \cite{berry}: it is
the connection of a Dirac monopole\footnote{As pointed out in
\cite{gphases}, the first appearance of the Dirac monopole was
actually in the context of the Born-Oppenheimer approximation for
diatomic systems, in what is now recognised as a Berry connection.
This was some two years before Dirac's work and more than fifty
years prior to Berry. The monopole tourist can view this
connection as the $\cos\theta$ term in equation (15) of
\cite{vanvleck}.}, with field strength
\be \vec{\nabla}\times \vec{A} = \frac{\vec{B}}{2
B^3}\label{dirac}\ee
The curvature singularity at $\vec{B}=0$ reflects the fact that
the spin-up and spin-down states become degenerate at this point.
Indeed, the Dirac monopole provides a good approximation to
Berry's phase in the vicinity of any two-state level crossing.
However, in more complicated quantum mechanical systems, far from
the degenerate point, it is typically difficult to compute the
Berry connection since exact expressions for the ground states
appearing in \eqn{time} are rarely known. In this paper we shall
show that the Berry connection is exactly computable in quantum
mechanical systems with ${\cal N}=(2,2)$ supersymmetry.

\para
The parameters that we will focus on are a triplet of masses
$\vec{m}=(m_1,m_2,m_3)$ that exist in supersymmetric quantum
mechanics. ${\cal N}=(2,2)$ supersymmetry can be thought of as the
dimensional reduction of ${\cal N}=1$ supersymmetry in four
dimensions, and the masses $\vec{m}$ arise as background values
for the spatial components of the four-dimensional gauge
field\footnote{The parameters $\vec{m}$ would not respect Lorentz
invariance in four-dimensional theories, which is the reason they
are perhaps less familiar than holomorphic parameters which
appear in the superpotential. The triplet of masses in quantum
mechanics is cousin to the real mass in three dimensions
\cite{ahiss} and the complex twisted mass in two dimensions
\cite{hh}.}. The fact that the masses $\vec{m}$ parameterize ${\bf
R}^3$, just like the magnetic fields of the simple Hamiltonian
\eqn{1st}, is no coincidence: the Berry connections that we will
find will be variants on the theme of the monopole.

\para
The paper is organized as follows. In Section 2, we describe some
general properties of ${\cal N}=(2,2)$ quantum mechanics,
including the multiplet structure, Lagrangians, symmetries and the
space of parameters. Sections 3 and 4 contain the  computations of
Berry's phase. Section 3 deals with systems with a single ground
state. We present a non-renormalization theorem, previously
derived by Denef \cite{denef}, which protects Berry's phase from
receiving perturbative corrections. This allows us to find exact
expressions for Berry's phase in complicated systems, far from
level crossing points. In each case, the Berry connection is
simply a sum of Dirac monopoles.

\para
In Section 4 we turn to the more interesting situation with $N>1$
ground states and study the corresponding  $U(N)$ Berry connection
\cite{wz}. The Witten index of supersymmetric systems \cite{index}
provides a natural mechanism to ensure the degeneracy of ground
states over the full range of parameters. This mechanism is
qualitatively different from the coset construction  \cite{wz} or
Kramers degeneracy \cite{avron} that has previously been used in
the study of non-Abelian Berry's phase and, in recent years, has
found application in condensed matter systems
\cite{bigzhang,middlezhang,smallzhang} and quantum computing
\cite{qc,qc2}. Our focus will be on the simplest supersymmetric
system admitting two ground states: the ${\bf CP}^1$ sigma-model
with potential governed by $\vec{m}\in {\bf R}^3$. We show that
BPS instantons carry the right fermi zero-mode structure to
contribute to the off-diagonal components of the non-Abelian Berry
connection, and we perform the explicit one-instanton calculation.
One of the interesting features of this calculation is that,
despite supersymmetry, the non-zero modes around the background of
the instanton do not cancel. We show that the Berry connection for
the ${\bf CP}^1$ sigma-model is the $SU(2)$ 't Hooft-Polyakov
monopole which, at large distances, looks like a Dirac monopole,
but with the singularity at the origin resolved by instanton
effects.

\para
The literature contains a few earlier discussions of the
relationship between Berry's phase and supersymmetry. The
$tt^\star$ equations of \cite{ttstar1,ttstar2,ttstar3} apply in
the context of ${\cal N}=(2,2)$ quantum mechanics, and deal with
the Berry connection as the complex parameters of the
superpotential are varied. This is in contrast to the present
paper where we vary the triplet of vector multiplet parameters.
The difference is somewhat analogous to the distinction between
the Coulomb branch and Higgs branch in higher-dimensional theories
and we shall make this analogy more complete in Section
\ref{param}. A discussion of Berry's phase that is more closely
related to the present paper appeared in the context of the matrix
theory description of a D2-brane moving in the background of a
D4-brane \cite{berk}. The calculation of \cite{berk} is
essentially identical to that of Section \ref{freesec}. In a
companion paper \cite{adsberry} we will describe a somewhat
different non-Abelian Berry's phase that occurs for a D0-brane
moving in the background of a D4-brane.

\para
{\bf Note Added:} Several aspects discussed in this paper have been 
clarified in later work. In \cite{bpsberry}, we showed that the 
the exact non-Abelian Berry connection discussed in Section 4 is the 
BPS 't Hooft-Polyakov monopole. Moreover, in \cite{holonomy} we showed 
that the Berry phase that arises from varying vector multiplet parameters 
always solves the Bogomolnyi monopole equations.

\section{Supersymmetric Quantum Mechanics}

Quantum mechanics with ${\cal N}=(2,2)$ supersymmetry follows from
the dimensional reduction of ${\cal N}=1$ supersymmetric theories
in four dimensions. The superalgebra has four real supercharges
and is sometimes referred to as $N=4A$ supersymmetry\footnote{In
contrast, $N=4B$ supersymmetry descends from ${\cal N}=(0,4)$
theories in two dimensions.}. The supercharges form a complex
doublet $Q_\alpha = (Q_-,Q_+)$ with the supersymmetry algebra
given by,
\be \{Q_\alpha,Q_\beta\}= \{\bar{Q}_\alpha,\bar{Q}_\beta\} = 0\ \
\ ,\ \ \ \{Q_\alpha,\bar{Q}_\beta\} = 2H\delta_{\alpha\beta} +
2\vec{\sigma}_{\alpha\beta}\cdot \vec{Z}\ee
The triplet of central terms $\vec{Z}$ can be thought of as the
momentum in the three reduced dimensions. The automorphism group
of the algebra is
\be R=SU(2)_R\times U(1)_R\ ,\label{r}\ee
under which the supercharges $Q$ transform in the ${\bf
2}_{\,+\!1}$ representation, while the central charges $\vec{Z}$
transform as ${\bf 3}_{\,0}$. We start with a brief review of the
different supersymmetric multiplets that we will make use of; for
the most part these are familiar from other theories with four
supercharges.

\subsubsection*{\it Vector and Linear Multiplets}

The Abelian vector multiplet $V$ contains a single gauge field
$A_0$. In quantum mechanics its role is to impose the constraint
of Gauss' law on the Hilbert space. The propagating degrees of
freedom consist of three real scalars $\vec{X}$ that arise from
the dimensional reduction of the four-dimensional gauge field, and
a pair of complex fermions $\lambda_\pm$. There is also an
auxiliary field $D$. Under the R-symmetry, $\vec{X}$ transforms in
${\bf 3}_{\,0}$ while $\lambda$ transforms in ${\bf 2}_{\,+1}$. We
will usually write $\lambda=(\lambda_-,\lambda_+)^T$.

\para
In higher-dimensional theories,  it is useful to consider the
gauge-invariant multiplet, which contains the Abelian field
strength as opposed to the gauge potential\footnote{For example,
in four dimensions the field strength is contained in the chiral
multiplet $W_\alpha= \bar{D}\bar{D}D_\alpha V$; in three
dimensions one may dualize the Abelian vector multiplet for a
linear multiplet defined by
$J=\epsilon^{\alpha\beta}\bar{D}_\alpha D_\beta V$; while in two
dimensions the relevant object is the twisted chiral multiplet
$\Sigma = \bar{D}_+ D_-V$.}. In quantum mechanics, the analogous
object is a triplet of linear multiplets $\vec{\Sigma}$
\cite{is,diac},
\be \vec{\Sigma} = \frac{1}{2}\bar{D}\vec{\sigma}D V  =  - \vec{X}
+i{\theta}\vec{\sigma}\bar{\lambda} + i
\bar{\theta}\vec{\sigma}\lambda - \bar{\theta}\vec{\sigma}\theta D
+\bar{\theta}(\vec{\sigma}\times\, \dot{\!\!\vec{X}}\,)\theta  +
\ldots
\nn\ee
The kinetic terms for the vector multiplet are given by%
\be L_{vector} = \frac{1}{g^2}\int d^2\theta d^2\bar{\theta} \
\vec{\Sigma}^2 = \frac{1}{2g^2}\ \dot{\!\!\vec{X}}{}^2 +
\frac{i}{g^2}\bar{\lambda}\dot{\lambda} + \frac{1}{2g^2}D^2 \ee
In non-relativistic quantum mechanics, the coefficient $1/g^2$
should be thought of as the mass of a particle moving in the
$\vec{X}$ direction; nonetheless we continue to use the notation
of coupling constants more appropriate to higher dimensional field
theories.

\subsubsection*{\it Chiral Multiplets}

The chiral multiplet $\Phi$ is familiar from four dimensions and
so we will be brief.  It  contains a complex scalar $\phi$ and a
pair of complex Grassmann fields $\psi_\pm$ which we again write
as $\psi=(\psi_-,\psi_+)^T$. There is also the complex auxiliary
field $F$. The scalars $\phi$ are invariant under $SU(2)_R$, while
the fermions are doublets. For a chiral multiplet with charge $q$
under the $U(1)$ vector multiplet, the Lagrangian is given by
\be L_{\rm chiral} &=& \int d^2\theta d^2\bar{\theta}\
\bar{\Phi}\,e^{2qV}\Phi \label{fullon} \\ &=&  | {\cal D}_t\phi|^2
+ \bar{\psi}(i{\cal D}_t-\vec{X}\cdot\vec{\sigma})\psi + |F|^2 +
qD|\phi|^2 -q^2\vec{X}^2|\phi|^2-i\sqrt{2}q
(\bar{\phi}\epsilon_{\alpha\beta}\psi_\alpha\lambda_\beta - {\rm
h.c.}) \nn\ee
%
with ${\cal D}_t\phi= \dot{\phi}-iqA_0\phi$. In this paper we will
work with Abelian gauged linear sigma models \cite{phases}, built
from a single $U(1)$ vector multiplet coupled to some number of
chiral multiplets.

\subsection{The Parameter Space}
\label{param}

The Berry connection is a gauge connection over the space of
parameters ${\cal M}$ of the theory. Supersymmetric quantum
mechanics has a number of different parameters, which can
be considered as background fields living in different
supermultiplets.

\para
As is familiar from many contexts, complex parameters that appear
in the superpotential lie in background chiral multiplets. Just as
in higher-dimensional field theories \cite{sholo}, certain
properties of the quantum mechanics depend analytically on these
parameters. For example, this holomorphic dependence is behind the
$tt^\star$ equations of \cite{ttstar1}. We can introduce a complex
mass parameter $\mu$ of this type only if we have two chiral
multiplets, $\Phi$ and $\tilde{\Phi}$, carrying opposite gauge
charge. We can then write the gauge-invariant superpotential,
\be {\cal W} = \mu\,\tilde{\Phi}\Phi\label{supper}\ee
A second class of parameters lives in the linear multiplets
$\vec{\Sigma}$. These are the triplet of mass parameters described
in the introduction. They are associated to weakly gauging a
$U(1)_F$ flavour symmetry of the quantum mechanics and, unlike the
complex mass parameter $\mu$, can be assigned to a single chiral
multiplet $\Phi$,
\be L_{\rm mass} = \int d^4\theta\, \Phi^\dagger
\exp\left(\bar{\theta}\,\vec{m}\cdot\vec{\sigma}\,\theta\right)\Phi
= |\dot{\phi}|^2 + i\bar{\psi}\dot{\psi} - m^2\,|\phi|^2 -
\bar{\psi}(\vec{m}\cdot\vec{\sigma})\psi \label{simple}\ee
Here $m=|\vec{m}|$. If $\Phi$ also carries charge $q$ under the
$U(1)$ gauge multiplet, the mass terms are given by
$(q\vec{X}-\vec{m})^2|\phi|^2$, with similar expressions for the
fermions.

\para
An important feature of these parameters is that it may not be
consistent with supersymmetry to turn on $\mu\neq 0$ and
$\vec{m}\neq 0$ at the same time. This can be most simply seen by
viewing $\vec{m}$ and $\mu$ as dynamical background
supermultiplets. To illustrate this, consider the theory with two
chiral multiplets $\Phi$ and $\tilde{\Phi}$ of gauge charge $+1$
and $-1$ respectively. One may introduce a triplet of masses
$\vec{m}$ by weakly gauging the $U(1)_F$ global symmetry under
which both $\Phi$ and $\tilde{\Phi}$ have charge $+1$. This will
give rise to masses,
\be (\vec{X}-\vec{m})^2|\phi|^2 +
(\vec{X}+\vec{m})^2|\tilde{\phi}|^2\label{trouble}\ee
\EPSFIGURE{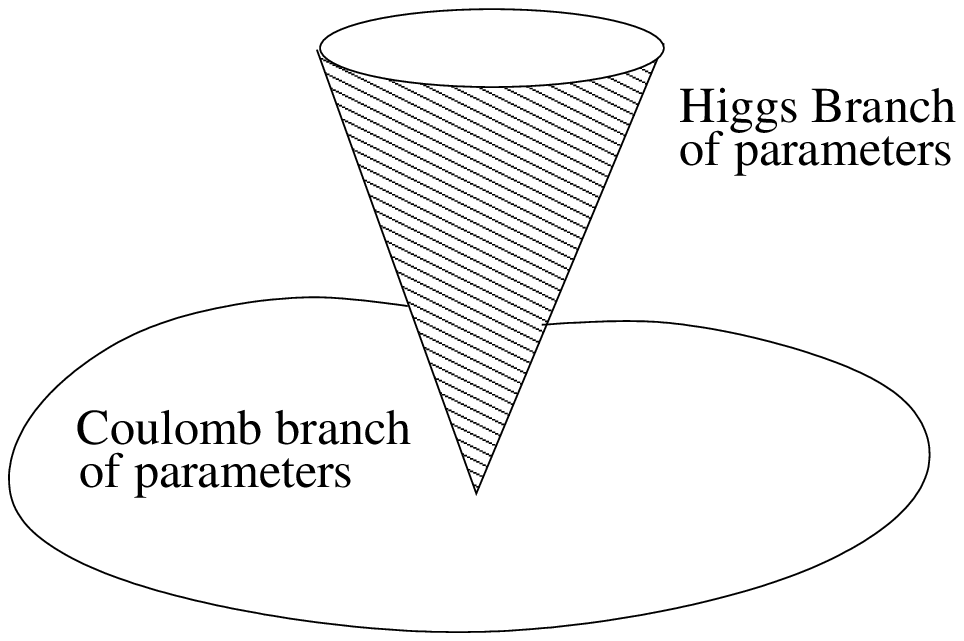,height=115pt}{The parameter space of the
supersymmetric quantum mechanics.}
\noindent However, invariance of the superpotential \eqn{supper}
under $U(1)_F$ requires that $\mu$ carries charge $-2$. This
results in the further contribution to the potential energy
$4m^2\,|\mu|^2$ which gives a non-zero ground-state energy, and
hence breaks supersymmetry, if both $\vec{m}$ and $\mu$ are
non-vanishing\footnote{Another way to see this is that the fermion
$\zeta$ which sits in the background chiral multiplet with $\mu$
has a non-zero transformation under supersymmetry, $\delta\zeta =
i\bar{\epsilon}(\vec{m}\cdot\vec{\sigma})\mu$, and cannot be set
to zero.}. This kind of behaviour is very familiar for dynamical
fields where it gives rise to the usual distinction between the
Coulomb branch and Higgs branch of vacua. Here we see the same
phenomenon at play in the space of parameters, rather than the
space of vacua. We use the same nomenclature. The space of
parameters of the theory is depicted in the figure: $\mu$ provides
a coordinate for the Higgs branch of parameters, while
$\vec{m}$ provide coordinates on the Coulomb branch. In this
paper, we will focus on the Berry connection over the Coulomb
branch of parameters.

\para
There are two further parameters of interest. We have already seen
the gauge coupling constant $g^2$ in the vector multiplet
Lagrangian. We may also introduce a real Fayet Iliopoulos (FI)
parameter $r$ by the $SU(2)_R$ invariant integral,
\be L_{FI} = r\int d\bar{\theta}\, \vec{\sigma}\, d\theta \cdot
\vec{\Sigma}  = rD\ee

\section{Abelian Berry's Phase}

In this section we discuss Berry's phase in systems with a single
ground state. We start with the simplest occurrence of Berry's
phase: a free chiral multiplet $\Phi$ with the mass triplet
$\vec{m}$. Since this will provide the  ``tree-level"
approximation to Berry's phase in more complicated systems, we
spend some time describing this basic set-up from both the
Hamiltonian and the Lagrangian viewpoints. The latter will provide
a useful spring-board to discuss Berry's phase in interacting
theories.

\subsection{Berry's Phase in a Free Theory}
\label{freesec}

Using the spinor notation $\psi^T=(\psi_-,\psi_+)$, the Lagrangian
for a free chiral multiplet $\Phi$ with mass $\vec{m}$ is,
\be  L_{\rm free} = |\dot{\phi}|^2 + i\bar{\psi}\dot{\psi} -
m^2\,|\phi|^2 - \bar{\psi}(\vec{m}\cdot\vec{\sigma})\psi\ .
\label{free}\ee
Note that the fermion mass term is reminiscent of the simple
Hamiltonian \eqn{1st} described in the introduction. This term
will indeed be responsible for Berry's phase.

\para
We pass to the Hamiltonian formalism by introducing the canonical
momenta $\pi=\dot{\phi}^\dagger$ and the Fermionic conjugate
momenta $\partial L/\partial \dot{\psi}=-i\bar{\psi}$, giving us a
Hamiltonian consisting of a two bosonic and two fermionic harmonic
oscillators,
\be H= |\pi|^2 +\vec{m}^2|\phi|^2 +
\bar{\psi}(\vec{m}\cdot\vec{\sigma})\psi \ee
We now quantize this theory using the canonical approach.  We use
the usual Schr\"odinger representation for the bosonic fields,
\be [\phi,\pi]= i \ \ \ \Rightarrow\ \ \  \pi =
-i\frac{\partial}{\partial \phi}\ee
while the fermionic anti-commutation relations
$\{\psi,\bar{\psi}\}=1$ are implemented by defining a reference
state $|\,0\ket$ annihilated by $\psi$: $\psi_+|\,0\ket =
\psi_-|\,0\ket=0$. We then form a basis of the fermionic Hilbert
space  by acting on $|\,0\rangle$ with the creation operators
$\bar{\psi}_{\pm}$ to form the four states,
\be |\,0\ket\ \ ,\ \ \bar{\psi}_+|\,0\ket\ \ \,\ \
\bar{\psi}_-|\,0\ket\ \ ,\ \ \bar{\psi}_+\bar{\psi}_-|\,0\ket\ee
We focus on the fermionic part of the Hamiltonian,
\be H_F=\bar{\psi}(\vec{m}\cdot\vec{\sigma})\psi \label{fham}\ee
The top and bottom states are both excited states in the Hilbert
space: they have $ H_F|\,0\rangle =
H_F\,\bar{\psi}_+\bar{\psi}_-|\,0\rangle = 0$ so that, once
dressed with the ground state of the complex boson $\phi$, they
have energy $+m$. Here we focus on the ground state of the system
which  lies in the two-dimensional fermionic Hilbert space ${\cal
H}_2$ spanned by $\bar{\psi}_\pm |\,0\rangle$. The fermionic
Hamiltonian \eqn{fham} acts on this space as,
\be H_F=\vec{m}\cdot\vec{\sigma}\label{eastham}\ee
%
%
$H_F$ has eigenvalues $\pm m$. The fermionic ground state
$|\,\Omega\rangle$ satisfies
$H_F|\,\Omega\rangle=-m|\,\Omega\rangle$ so that, once dressed
with the bosonic ground state, it yields a state with vanishing
energy as expected in a supersymmetric system.

\para
The Hamiltonian \eqn{eastham} acting on the ground state coincides
with that of a spin 1/2 particle in a magnetic field \eqn{1st}.
Correspondingly, the Berry connection of the ground state as the
parameters $\vec{m}$ are varied is given by the Dirac monopole. To
give an explicit form of the connection, we should first pick a
gauge which, in this context, means reference choice of ground
state --- including the phase --- for each $\vec{m}$. To this end,
we define the projection operator onto the ground state
\be P_{-} = \frac{1}{2}\left({\bf 1} -
\frac{\vec{m}}{m}\cdot\vec{\sigma} \right) \ee
Then we define the phase of our reference ground states to be that
of the (un-normalized) state $P_-\,\bar{\psi}_+|\,0\rangle$, which
is valid everywhere except along the half-line $\vec{m}=(0,0,-m)$
where the ground state $|\Omega\rangle$ is orthogonal to
$\bar{\psi}_+|\,0\rangle$. As a result, the Berry connection has a
Dirac string singularity along this axis. The Berry connection is
$\vec{A}=i\langle\Omega|\vec{\nabla}|\Omega\rangle$.
%
%
It is a simple matter to compute the explicit connection which is
given by the Dirac monopole. In Cartesian coordinates, it is
\be
 A^{\rm Dirac}_i = \frac{\epsilon_{ij}m_j}{2m(m+m_3)}\ \ \ {i,j=1,2}\ \ \ {\rm
 and}\ \ \ A^{\rm Dirac}_3=0\label{berry}\ee
 %
 %
%
For a closed, adiabatic variation of the parameters $\vec{m}$, the
Berry phase is the then given by the integral of this connection so
that
\be |\,\Omega\rangle \rightarrow  \exp\left(-i\oint \vec{A}\cdot
d\vec{m}\right) |\,\Omega\rangle\ee

\subsection{The Born-Oppenheimer Approximation}

One man's fixed parameter is another's dynamical degree of
freedom. This is the essence of the Born-Oppenheimer approximation
in which the ``parameters" of the model are not really fixed, but
merely slowly moving degrees of freedom. We may endow the
parameters $\vec{m}$ with dynamics by introducing the canonical
kinetic terms
\be L_{m} = \frac{1}{2e^2}\,\dot{\vec{m}}{}^2\label{mkin}\ee
The Born-Oppenheimer approximation is then valid if $e^2 \ll
\langle m\rangle^3$. This ensures that the $\phi$ fields have high
frequency and may be treated in a fixed $\vec{m}$ background. From
a modern perspective, this is equivalent to the Wilsonian approach
to quantum mechanics, in which the fast moving $\phi$ degrees of
freedom are integrated out. (Note, however, that in quantum
mechanics the fast moving degrees of freedom are the physical
light particles, while in field theory they are the virtual heavy
particles). As in field theory, the Born-Oppenheimer-Wilson
framework is ideally suited to working in the language of path
integrals and effective actions rather than Hamiltonians
\cite{japanese,moody}. In this section we show how to reproduce
the Berry's phase from this perspective.

\para
We wish to integrate out the $\phi$ and $\psi$ fields in
\eqn{free} in the background of time varying parameters
$\vec{m}(t)$. This results in contributions to the effective
action for $\vec{m}$. The contribution from the bosons $\phi$ is
given by
\be {\cal L}_{\rm bose} = \log\det\left(\frac{-\partial_t^2 -
m(t)^2}{-\partial_t^2-{m}_0^2}\right)\ee
where the vacuum value for the masses may be taken to be
$\vec{m}_0=\frac{1}{T}\int_0^T dt\,\vec{m}(t)$. The determinant
may be easily computed as an expansion of $\dot{\vec{m}}$. The
first two terms are given by
\be {\cal L}_{\rm bose} = -m +
\frac{\dot{\vec{m}}{}^2}{8m^3}+\ldots \label{lbose}\ee
Here the first term corresponds to the usual zero-point energy of
$\phi$, while the second term can be interpreted as a finite
renormalization of the kinetic term $1/e^2 \rightarrow 1/e^2 +
1/4m^3$.

\para
The contribution from the fermions $\psi$ is given by the
determinants
\be {\cal L}_{\rm fermi} =
-\log\det\left(\frac{i\partial_t-\vec{m}(t)\cdot\vec{\sigma}}{i\partial_t
- \vec{m}_0\cdot\vec{\sigma}}\right)\ee
These determinants include the Berry phase, as first shown by
Stone \cite{stone}. The simplest way to see this is to recall that
the determinants compute the vacuum-vacuum amplitude. To leading
order, this includes the dynamical phase $e^{i\int dt\, m(t)}$ and
the Berry phase $e^{i\int dt\,\vec{A}\cdot \dot{\vec{m}}}$. This
translates into an effective action,
\be {\cal L}_{\rm fermi} = +m + \vec{A}(\vec{m})\cdot
\dot{\vec{m}}+ \ldots\ee
The first term is the zero point energy of a Grassmannian variable
and cancels the contribution in \eqn{lbose} as expected in a
supersymmetric theory. The Berry connection appearing in the
second term is the Dirac monopole \eqn{berry}. There is no further
renormalization to the kinetic terms from the fermions. In
summary, the effect of these simple one-loop computations  in
quantum mechanics is to provide an  effective  low-energy dynamics
of the parameters $\vec{m}$ which, up to two derivatives, is given
by
\be  L_{\rm eff} =  \vec{A}\cdot
\dot{\vec{m}}+\left(\frac{1}{2e^2}+\frac{1}{8m^3}\right)\dot{\vec{m}}{}^2+\ldots
\label{berrygood}\ee
with the first term identified as the Berry connection.

\subsubsection*{The Supersymmetric Completion}

As we reviewed in Section \ref{param}, the parameters $\vec{m}$
can be thought of as living in a supersymmetric vector multiplet
$V$. As well as the three parameters $\vec{m}$, this multiplet
also contains a gauge field $u$, an auxiliary scalar $D$ and two
complex fermions $\eta_{\pm}$. If we are to apply the
Born-Oppenheimer approximation in a supersymmetric manner, the
kinetic terms \eqn{mkin} must be accompanied with suitable terms
for the other fields in the multiplet, together with further
Yukawa coupling interactions. For $e^2\neq 0$, the Lagrangian
\eqn{mkin} should be replaced by,
\be L_{m} &=& \frac{1}{2e^2}\dot{\vec{m}}{}^2 +
\frac{i}{e^2}\bar{\eta}\dot{\eta}+\frac{1}{2e^2}D^2\label{mkin2}\ee
while \eqn{free} is now generalized to include interaction terms
between the chiral multiplet and vector multiplet fields, given by
\eqn{fullon} with $\vec{X}$ replaced by the parameter $\vec{m}$,
and $\lambda$ replaced by $\eta$.
%
%
In the Born-Oppenheimer approximation, we once again should
integrate out the chiral multiplet in the regime $e^2 \ll \langle
m\rangle^3$. We now search for an effective action for the vector
multiplet that itself preserves supersymmetry. Such an action was
derived by Denef in \cite{denef}. (As action of the same form was
also derived previously by Smilga in the study of the zero mode
dynamics of SQED \cite{smilga,bsmilga}. Integrating out the chiral
multiplets gives the supersymmetric completion of the Berry term,
\be {\cal L}_{\rm Berry} = \vec{A}\cdot\dot{\vec{m}} -\frac{1}{2m}
D + \bar{\eta}\,\frac{\vec{m}\cdot\vec{\sigma}}{2m^3}\eta
\label{stillgood}\ee
This Lagrangian is invariant under the supersymmetry
transformations,
\be \delta u &=& i\bar{\eta}\xi - i\bar{\xi}\eta \nn\\ \delta
\vec{m} &=& i\bar{\eta}\vec{\sigma}\xi - i \bar{\xi}\vec{\sigma}
\eta \label{transformers}\\ \delta \eta &=&
\dot{\vec{m}}\cdot\vec{\sigma}\xi + iD \xi \nn\\ \delta D &=&
-\dot{\bar{\eta}}\xi - \bar{\xi}\dot{\eta}\nn \ee
One of the consequences of supersymmetry is that the presence of
kinetic terms for parameters also introduces new interaction
terms. These are written in equation \eqn{fullon} and they further
affect the dynamics of the theory. For example, the D-term
interactions in \eqn{fullon} give rise to a one-loop potential
over the parameter space. This is then captured in the effective
theory \eqn{stillgood} by the D-term which gives rise to the
potential,
\be V=
\frac{1}{8m^2}\left(\frac{1}{e^2}+\frac{1}{4m^3}\right)^{-1}\ee
where we have invoked the coupling renormalization
\eqn{berrygood}. Thus the true supersymmetric vacuum lies at
$m\rightarrow \infty$. One may wonder whether these interactions
also change the Berry connection, which was computed in the strict
$e^2=0$ limit. In principle there could be $e^2/m^3$ corrections
to the connection. The fact that this cannot happen follows from a
non-renormalization theorem proven by Denef \cite{denef}. He
considered the most general Lagrangian containing a single time
derivative and fermi bi-linear terms, consistent with the
$SU(2)_R$ symmetry of the model,
\be {\cal L} = \vec{A}(\vec{m})\cdot\dot{\vec{m}} - U(m)D +
C(m)\bar{\eta}\eta +
\bar{\eta}\,\vec{C}(\vec{m})\cdot\vec{\sigma}\eta\label{general}\ee
with arbitrary functions $\vec{A}$, $U$, $C$ and $\vec{C}$.
Requiring that this action is supersymmetric places strong
constraints on these functions. The supersymmetry transformations
\eqn{transformers} are dictated (at this order) by the superspace
formulation \cite{diac}. Invariance of the Lagrangian then
requires,
\be \vec{C}={\nabla}U = {\nabla}\times \vec{A}\ \ \ \ {\rm and}\ \
\ \ C=0\label{restrict}\ee
Allowing for singular behaviour at the origin, the most general
spherically symmetric solution to these constraints is the Dirac
monopole connection \eqn{berry}. This is the promised
non-renormalization theorem for the Berry connection. We will now
apply this to compute Berry's phase in more complicated,
interacting theories.

\subsection{Berry's Phase in Interacting Theories}
\label{int}

We now turn to interacting theories where more complicated Berry's
connections may be expected. In this section we restrict to
theories with a single ground state, ensuring an Abelian Berry
connection. We will treat the more interesting theories with
multiple, degenerate ground states in the following section.

\para
We consider the $U(1)$ gauge theory with two chiral multiplets
$\Phi$ and $\tilde{\Phi}$ of charge $+1$ and $-1$ respectively.
This is the dimensional reduction of ${\cal N}=1$ SQED. The
bosonic part of the Lagrangian is given by,
\be L=\frac{1}{2g^2}\dot{\vec{X}}{}^2 + |{\cal D}_t\phi|^2 +
|{\cal D}_t\tilde{\phi}|^2 - X^2(|\phi|^2+|\tilde{\phi}|^2) -
\frac{g^2}{2}(|\phi|^2-|\tilde{\phi}|^2-r)^2\label{boslag}\ee
Note that we have introduced a FI parameter $r>0$. Although the
vector multiplet scalars $\vec{X}$ look ripe to treat in the
Born-Oppenheimer approximation, we will not do so here; instead we
take the  limit $g^2 \rightarrow \infty$ where the theory reduces
to a gauged linear sigma-model with target space defined by the
D-flatness conditions,
\be |\phi|^2-|\tilde{\phi}|^2=r\label{d}\ee
modulo the gauge action $\phi\rightarrow e^{i\alpha}\phi$ and
$\tilde{\phi}\rightarrow e^{-i\alpha}\tilde{\phi}$. Asymptotically
this is the cone ${\bf C}/{\bf Z}_2$. The conical singularity is
resolved by the FI parameter $r>0$. Since we are not treating
$\vec{X}$ as the parameters for Berry's connection, we must
introduce different parameters. We will do that now.

\subsubsection*{The Parameter Space}

We wish to introduce a potential on the target space. As discussed
in Section \ref{param}, there is a Higgs branch and Coulomb branch
of parameters for this model. The physics on these two branches is
very different.

\para
Moving on the Higgs branch of parameters requires us to introduce
the gauge invariant superpotential ${\cal W} =
\mu\,\tilde{\Phi}\Phi$. When $\mu\neq 0$, there is no simultaneous
solution to the D-term constraints \eqn{d} and the F-term
constraints $\phi=\tilde{\phi}=0$, and supersymmetry is
spontaneously broken.

\para
Moving on the Coulomb branch of parameters introduces the triplet
of masses $\vec{m}$ arising from weakly gauging the $U(1)_F$
flavour symmetry of the model, under which $\Phi\rightarrow
e^{i\alpha}\Phi$ and $\tilde{\Phi}\rightarrow
e^{i\alpha}\tilde{\Phi}$. The masses sit in the Lagrangian by
replacing the $X^2(|\phi|^2+|\tilde{\phi}|^2)$ term in
\eqn{boslag} by \eqn{trouble}.
%
%
In contrast to the complex mass, the presence of $\vec{m}$ does
not break supersymmetry. The theory has a unique classical ground
state given by
\be \vec{X}=\vec{m}\ \ \ ,\ \ \ |\phi|^2 = r\ \ \ ,\ \
\tilde{\phi}=0 \ee
This behaviour is consistent with the Witten index because the
point $\mu = \vec{m} =0$ is singular: here the potential on the
non-compact vacuum moduli space vanishes, allowing  the ground
state wavefunction to spread into the asymptotic regime of ${\bf
C}/{\bf Z}_2$ and become non-normalizable. At this point, the
zero-energy state exits the Hilbert space of the theory and the
theory breaks supersymmetry.

\subsubsection*{Berry's Phase}

On the Coulomb branch of parameters, the dimensionless coupling
for our massive sigma model is $1/mr$. For $mr \gg 1$, the ground
state wavefunction is restricted to a region of field space much
smaller than the curvature of the target space, and is well
approximated by the free theory described in Section
\ref{freesec}. The Berry connection for the ground state is, to
leading order, described by the Dirac monopole \eqn{berry}.
However, we naively expect corrections to the Berry connection,
which can be computed perturbatively in $1/mr$, as the ground
state wavefunction begins to feel the curvature of the target
space. The fact that such corrections cannot occur has nothing to
do with the supersymmetric non-renormalization theorem, but
instead follows from the fact that the quantized flux emitted from
the singularity is conserved as one moves in parameter space. This
ensures that any corrections to the curvature are solenoidal in
nature. Yet there are no such corrections consistent with the
$SO(3)$ rotational symmetry of the problem.


\subsubsection{Further Examples}

It is a simple matter to cook up related models with a single
ground state without $SO(3)$ symmetry. We could consider the
gauged linear sigma model built from $U(1)$ with a single chiral
multiplet $\Phi$ of charge $+1$ and $M$ chiral multiplets
$\tilde{\Phi}_i$ of charge $-1$. The D-term constraint is
\be |\phi|^2-\sum_{i=1}^M |\tilde{\phi}_i|^2 = r \ee
\EPSFIGURE{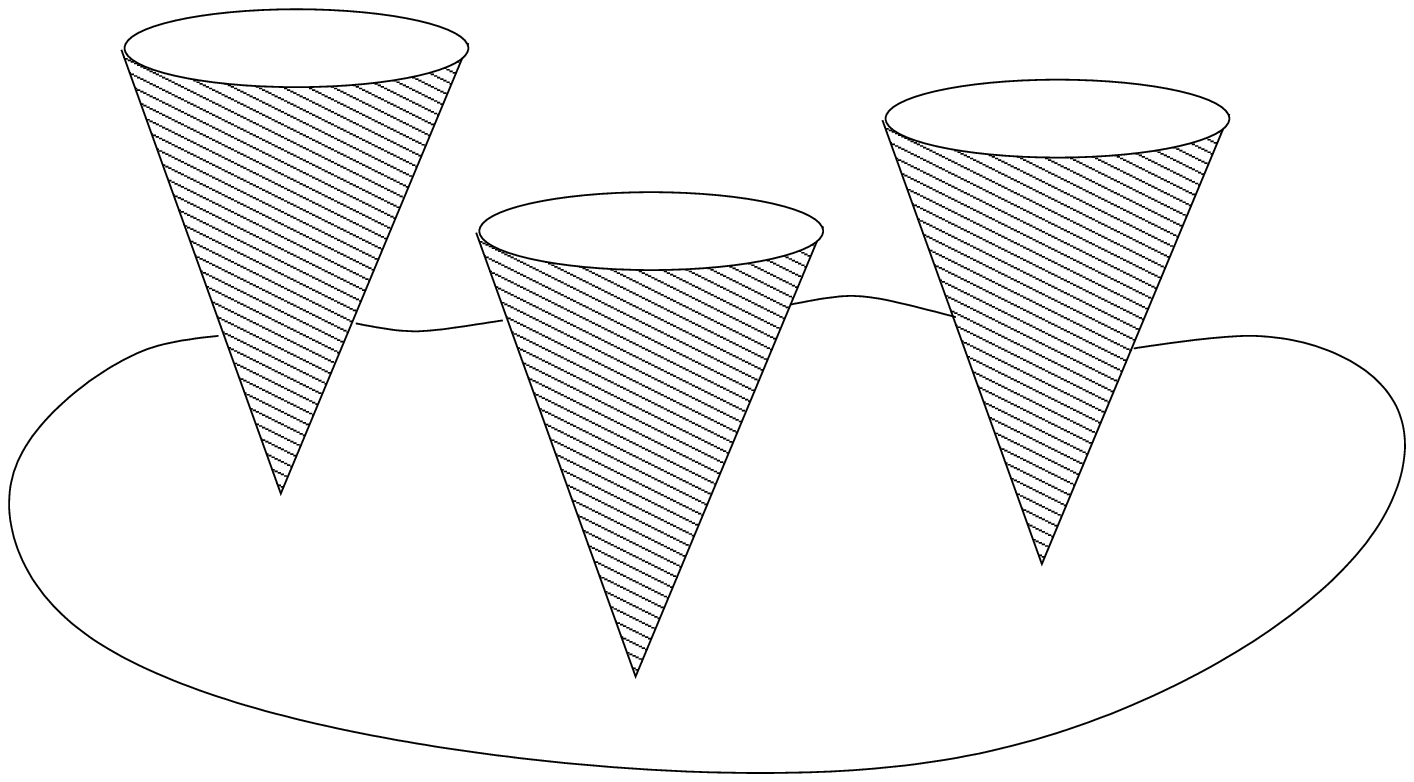,height=110pt}{The parameter space with
multiple chiral multiplets.}
\noindent We endow $\Phi$ with the triplet of masses $\vec{m}$,
and each $\tilde{\Phi}_i$ with masses $\vec{\tilde{m}}_i$. For $r
> 0$ and $\vec{m}\neq \vec{\tilde{m}}_i$, the theory has a unique
classical ground state given by $\vec{X}=\vec{m}$ and $|\phi|^2=r$
with $\tilde{\phi}_i=0$. For the present purposes we will fix
$\vec{\tilde{m}}_i$ and ask about the behaviour of the ground
state as we vary $\vec{m}$. We expect $M$ singular points, lying
at $\vec{m}=\vec{\tilde{m}}_i$, at which a branch of Higgs vacua
emerges. Alternatively, we can view these as points where a new
Higgs branch of supersymmetric parameters is admitted,
corresponding to a superpotential ${\cal
W}=\mu_i\tilde{\Phi}_i\Phi$, so that the parameter space of the
theory looks like that shown in Figure 2.

\para
Now the symmetries of the problem are not sufficient to rule out
general solenoidal contributions to the Berry connection. However,
the non-renormalization theorem of the previous section is: we may
integrate out all fields to derive an effective action for the
parameters $\vec{m}$ which is of the form \eqn{general}. The
constraints \eqn{restrict} are now satisfied by a superposition of
Dirac monopoles, lying at positions $\vec{m}=\vec{\tilde{m}_i}$,
\be \nabla\times \vec{A} = \sum_{i=1}^M
\frac{\vec{m}-\!\vec{\,\,\tilde{m}_i}}{2|\vec{m}-\!\vec{\,\,\tilde{m}_i}|^3}\ee
In summary, the restriction of supersymmetry requires that the
Berry's phase does not receive any corrections from its simple
``tree-level" value and, even in the full interacting theory is
still given by the Dirac monopole. While it is pleasing to be able
to make precise statements about objects in interacting quantum
mechanics it is, nonetheless, a little disappointing that the
object is not particularly novel. In the next section we will
instead turn to a situation where a more interesting Berry
connection emerges.

\section{Non-Abelian Berry's Phase}

Supersymmetry provides a natural arena in which to study
non-Abelian Berry's phases. Unlike previous examples
\cite{wz,avron}, the the degeneracy of ground states is dictated
not by a symmetry, but rather by an index. In this section we
discuss a simple supersymmetric system with two ground states. We
compute the $U(2)$ Berry connection over the ${\bf R}^3$ parameter
space. We show that it is given by a {\it smooth} 't
Hooft-Polyakov monopole.

\subsection{The ${\bf CP}^1$ Sigma Model}

Our simple model is  supersymmetric quantum mechanics with ${\bf
CP}^1$ target space and a potential. We construct this target
space from a gauged linear theory with $U(1)$ gauge group and two
chiral multiplets, both of charge $+1$. The bosonic part of the
Lagrangian is given by,
\be\ L=\frac{1}{2g^2}\dot{\vec{X}}{}^2 + \sum_{i=1}^2|{\cal
D}_t\phi_i|^2 - (\vec{X}-\vec{m}_i)^2|\phi_i|^2  -
\frac{g^2}{2}(|\phi_1|^2+|\phi_2|^2-r)^2\label{cp1}\ee
In the limit $g^2\rightarrow \infty$, the D-term restricts us to
${\bf S}^3$ defined by,
\be |\phi_1|^2+|\phi_2|^2  = r\ee
Further dividing by gauge transformations $\phi_i\rightarrow
e^{i\alpha}\phi_i$ leaves us with the target space ${\bf
CP}^1\cong {\bf S}^2$. The triplets of mass parameters $\vec{m}_i$
induce a potential on this space. By a suitable shift of $\vec{X}$
we may choose
\be \vec{m}_1=-\vec{m}_2\equiv\vec{m} \ee
We are interested in computing the Berry connection that arises as
we adiabatically vary $\vec{m}$. (The ${\bf CP}^1$ model was also
the system in which the $tt^\star$ equations were first applied to
compute Berry's phase, this time as the complexified K\"ahler
class $t=r+i\theta$ of the $d=1+1$ dimensional theory is varied
\cite{ttstar3}).

\para
In the absence of masses $\vec{m}$, the model admits an $SU(2)_F$
global symmetry, transforming $\phi_i$ in the doublet
representation. This is simply the isometry of the ${\bf CP}^1$
target space. The masses can be thought of as living in a
background vector multiplet for $SU(2)_F$ and therefore transform
in the adjoint representation. Since $\vec{m}$ also transforms in
the ${\bf 3}$ of $SU(2)_R$, it breaks the non-Abelian global
symmetries to their Cartan subalgebra\footnote{The subgroup
$U(1)_A\subset SU(2)_R$ may be identified with the axial $U(1)_A$
symmetry in two-dimensional ${\cal N}=(2,2)$ theories.}
\be \vec{m}: SU(2)_F\times SU(2)_R \rightarrow U(1)_F\times
U(1)_A\label{downdown}\ee
The global $U(1)_F$ symmetry acts on the bosons as
$\phi_1\rightarrow e^{i\theta}\phi_1$ and $\phi_2\rightarrow
e^{-i\theta}\phi_2$ and will play a role in the discussion of
instantons. For non-zero $\vec{m}$, the theory has two, isolated,
classical vacua. They are
\begin{itemize}
\item Vacuum 1: $\vec{X}=\vec{m}$ with $|\phi_1|^2 = r$ and
$\phi_2=0$. \item Vacuum 2: $\vec{X}=-\vec{m}$ with  $\phi_1 = 0$
and $|\phi_2|^2=r$.
\end{itemize}
The Witten index ensures that both of these vacua survive in the
quantum theory \cite{index}. This introduces yet another $U(2)_G$
symmetry, which is the gauge symmetry rotating these ground states
in the quantum theory. This is the reason that our Berry
connection is, a priori, a $U(2)$-valued object over ${\bf R}^3$.
However, in this case there exists a distinguished basis of ground
states, given by the eigenstates under $U(1)_F$. (In fact,
$U(1)_A$ would do equally well for these purposes). The fact that
the ground states carry different quantum numbers under
$U(1)_{F/A}$ leads to ``spontaneous breaking" of the $U(2)_G$
symmetry to the Cartan subalgebra, which can then be identified
with $U(1)_R\times U(1)_F$. Note that at the origin of parameter
space $\vec{m} = 0$, there is no breaking \eqn{downdown} and,
correspondingly, no way to distinguish the two ground states; here
$U(2)_G$ remains unbroken.

\subsubsection*{Berry's Phase}

The dimensionless coupling of our model is $1/mr$. In the limit
$mr\gg 1$, the physics around each vacuum is described by a
single, free chiral multiplet and, to leading order, the ground
state is simply Gaussian. In vacuum 1, $\phi_1$ is eaten by the
Higgs mechanism and the remaining dynamical field is $\Phi_2$,
with mass $+2\vec{m}$. As explained in Section \ref{freesec}, the
$U(1)$ Berry's connection for this vacuum is the Dirac monopole
connection $\vec{A}{}^{\rm Dirac}$. In contrast, in vacuum 2 the
free chiral multiplet is $\Phi_1$ with mass $-2\vec{m}$ and the
Berry connection is $-\vec{A}{}^{\rm Dirac}$. Putting these two
results together, the leading order $U(2)$ Berry connection is
given by
\be \vec{A} = \vec{A}{}^{\rm Dirac}\,\sigma^3\label{ddirac}\ee
which, in fact, lies in $SU(2)$ rather than $U(2)$.  This
connection inherits the Dirac string singularity along the
$\vec{m}=(0,0,-m)$ half-axis. However, as is well known, one may
eliminate this through the use of a  singular $SU(2)$ gauge
transformation,
\be U =
\exp\left(-\frac{i\theta(m_2\sigma^1-m_1\sigma^2)}{2\sqrt{m^2-m_3^2}}
\right)\label{singfromhigh}\ee
under which the Dirac monopole-anti-monopole pair \eqn{ddirac}
becomes $\tilde{A}_\mu= UA_\mu U^\dagger - i (\partial_\mu
U)U^\dagger$ which takes the rotationally symmetric form,
\be \tilde{A}_\mu =
\epsilon_{\mu\nu\rho}\,\frac{m_\nu}{2m^2}\sigma^\rho
\label{nickcave}\ee
As is clear in this gauge, the $U(2)_G$ gauge symmetry of the
ground states is locked with the $SU(2)_R$ symmetry rotating
$\vec{m}$.

\para
Equation \eqn{nickcave} is the leading order form of the Berry
connection, valid in the regime $mr \gg 1$. Like the Dirac
monopole, it is singular at $mr = 0$. In the Abelian case, this
singularity reflected the existence of a new ground state. Does a
similar phenomenon occur in the non-Abelian case? In fact, it
cannot. When $\vec{m}=0$, we have the usual ${\bf CP}^1$
sigma-model without potential. Witten showed many years ago that
the vacuum states of this model correspond to the
cohomology\footnote{In making the map to cohomology for a
non-linear sigma-model, one usually chooses the $SU(2)_R$
violating quantization condition
$\psi_{+i}|0\rangle=\bar{\psi}_{-i}|0\rangle=0$, with the map of
creation operators to differential forms:
$\bar{\psi}_{+i}\rightarrow d\bar{z}_i$ and $\psi_{-i}\rightarrow
(-1)^{N}dz^i$. Here we have instead quantized in a manifestly
$SU(2)_R$ invariant fashion, with
${\psi}_{i+}|0\rangle={\psi}_{i-}|0\rangle=0$. The quantum physics
remains unchanged, but the map to cohomology requires alteration.
This is the reason that the ground states of our model have a
single fermion excited, while the cohomology of ${\bf CP}^1$ is
even.} of ${\bf CP}^1$ \cite{index}: there are precisely two
ground states at $mr=0$. Since there is no extra degeneracy in the
ground state spectrum, there should be no singularity in the Berry
connection. We conclude that the Berry connection in the full
theory must take the form,
\be \tilde{A}_\mu =
\epsilon_{\mu\nu\rho}\,\frac{m_\nu}{2m^2}\sigma^\rho\left(1-f(mr)\right)
\label{interventionist}\ee
where the structure is fixed by $SU(2)$ covariance, and the
profile function has asymptotics
\be f(mr) \rightarrow \left\{\begin{array}{cl} 0\ \ \ \ \  & mr
\rightarrow \infty \\ 1+ {\cal O}(mr)^2\ \ \ \  & mr \rightarrow 0
\end{array}\right.\label{smooth}\ee
This is precisely the form of the 't Hooft-Polyakov monopole
\cite{t,p}. It remains to determine the function $f(mr)$. In the
previous Section we showed that there are no perturbative
corrections to the Berry connection. We will now show that the
relevant corrections are BPS instantons.

\subsection{Instantons}

In supersymmetric quantum field theories, the objects that receive
contributions from BPS instantons are typically rather special:
they are protected ``BPS" quantities that have been much studied
over the past decade or more. In theories with ${\cal N}=(1,1)$
supersymmetry, Witten famously showed that BPS instantons tunnel
between vacua with Morse index differing by one; they play a
crucial role in deriving the strong form of Morse inequalities
\cite{morse}. However, the question of which physical quantity
receives instanton corrections in quantum mechanics with extended
supersymmetry has not been satisfactorily answered. Here we show
that, in the case of ${\cal N}=(2,2)$ supersymmetry, the relevant
object is the Berry connection.

\subsubsection*{The Instanton Equations}

Kinks in the ${\bf CP}^1$ sigma model with potential were first
discussed in \cite{qkink}. To derive the first order equations
obeyed by the kinks, we first Wick rotate to Euclidean time $\tau
= -it$. For simplicity, we choose the masses to be aligned along
$\vec{m}=(0,0,m)$. The kink profile then takes the form
$\vec{X}=(0,0,X(\tau))$.  The bosonic part of the Euclidean action
can then be written as,
\be S_E &=& \int d\tau \ \frac{1}{2g^2}(\partial_\tau X)^2 +
\sum_{i=1}^2 |{\cal D}_\tau \phi_i|^2 + (X-m_i)^2|\phi_i|^2 +
\frac{g^2}{2}(|\phi_1|^2 + |\phi_2|^2 -r )^2\nn\\ &=& \int d\tau\
\frac{1}{2g^2}(\partial_\tau X - g^2(|\phi_1|^2 + |\phi_2|^2 -
r))^2 + (\partial_\tau X)(|\phi_1|^2+|\phi_2|^2-r) \nn\\
&& \ \ \ \ \ \ \ \ \ \ + \sum_i|{\cal D}_\tau \phi_i -
(X-m_i)\phi_i|^2 + ({\cal D}_\tau \phi_i\,\phi_i^\dagger (X-m_i) +
{\rm h.c.})\ee
Integrating the last term by parts partially cancels the other
cross term, and provides the Bogomolnyi bound for the kink,
\be S_{\rm kink} \geq r\int_{-\infty}^{+\infty} d\tau\
\partial_\tau X = 2mr \label{act} \ee
where the boundary conditions are chosen such that the kink
interpolates from vacuum 1 to vacuum 2 as $\tau$ increases, i.e.
\be X\rightarrow \left\{\begin{array}{cl} -m \ \ \ \ \ \ \ & \tau
\rightarrow +\infty \\ +m \ \ \ \ \ \ \ & \tau \rightarrow
-\infty\end{array}\right.\ee
The bound on the action \eqn{act} is saturated when the Bogomolnyi
equations for the kink are satisfied,
\be \partial_\tau X &=& g^2 (|\phi_1|^2+|\phi_2|^2-r)\nn\\ {\cal
D}_\tau\phi_i &=& (X-m_i)\phi_i\label{bog}\ee
While analytic solutions to these equations are not known for
general finite $g^2$, it is a simple matter to solve them in the
$g^2\rightarrow \infty$ limit (see, for example, \cite{me}) where
the $U(1)$ gauge theory reduces to the ${\bf CP}^1$ sigma-model.

\subsubsection*{Fermions}

The crux of the instanton calculation lies, as always, in the
fermion zero modes. After Wick rotation the equations of motion
for the fermions take the form
\be \Delta \left(\begin{array}{c}\lambda_-
\\ \bar{\psi}_{i+}\end{array}\right) =
\Delta^\dagger\left(\begin{array}{c}\lambda_+
\\ \bar{\psi}_{i-}\end{array}\right)  = 0
 \ee
where, in the vacuum $\vec{m}_i=(0,0,m_i)$ with the ansatz
$\vec{X}=(0,0,X)$, the Dirac operators take the form
\be \Delta =
 \left(\begin{array}{cc} \frac{1}{e^2}\,\partial_\tau
&
 -i\sqrt{2} \phi_i \\ i\sqrt{2}\bar{\phi}_i\ \ \ \  & -{\cal D}_\tau
 + (X-m_i) \end{array}\right),\ \
\Delta^\dagger = \left(\begin{array}{cc}
-\frac{1}{e^2}\,\partial_\tau &
 -i\sqrt{2} \phi_i \\ i\sqrt{2}\bar{\phi}_i\ \ \ \  & {\cal D}_\tau
 +
 (X-m_i) \end{array}\right)\ \ \ \ \label{dops}\ee
One can check that $\Delta$ has zero modes, while $\Delta^\dagger$
has none. (For example, $\Delta\Delta^\dagger$ is a positive
definite operator, with the off-diagonal components vanishing on
the instanton equations \eqn{bog}). This ensures that the kink has
two fermionic zero modes, carried by the pairs
$(\lambda_-,\bar{\psi}_{i+})$ and $(\bar{\lambda}_+,\psi_{i-})$.

\para
The existence of these fermi zero modes guarantees that the
instantons do not lead to tunnelling between vacuum states, but
instead contribute only to two-fermi correlation functions. We now
show that these correlation functions can be identified with the
Berry connection. With $\vec{m}= (0,0,m)$, the first vacuum state
is given by $|1\rangle= \bar{\psi}_{2+}|0\rangle$. From this, we
write the variation of this vacuum state as we change $\vec{m}$,
\be
\partial_{m_1}|1\rangle = -i\partial_{m_2}
|1\rangle=\frac{i}{2m}\bar{\psi}_{2+}\psi_{2-}|1\rangle\ \ \ ,\ \
\
\partial_{m_3}|1\rangle=0 \ee
where all derivatives are evaluated at $\vec{m}=(0,0,m)$. The
off-diagonal components of the non-Abelian Berry connection at
this point are therefore given by,
\be (A_{1})_{12} = -i(A_2)_{12} = -\frac{1}{2m}\,\langle 2|
\bar{\psi}_{2-}\psi_{2+}|1\rangle \ \ \ ,\ \ \  (A_3)_{12}=0
\label{thematrix}\ee
But, from the zero-mode analysis above, this matrix element
receives contributions from instantons. It is worth commenting on
the symmetries at this point. The ground states $|1\rangle$ and
$|2\rangle$ carry charge $+1$ and $-1$ respectively under
$U(1)_A$. This alone ensures that their overlap is vanishing
$\langle 1|2\rangle = 0$. However, the zero modes of the instanton
also carry $U(1)_A$ charge 2, so that the matrix element arising
in Berry's phase can be non-zero.

\para
We now compare this to the rotationally symmetric form of the
gauge connection \eqn{interventionist}. Performing the inverse
gauge transformation \eqn{singfromhigh}, we find that the
subsequent connection at the point $\vec{m}=(0,0,m)$ is
\be A_i=\frac{f}{2m}\epsilon_{ij}\,\sigma_j\ \ \ \ i,j=1,2 \ \ \
{\rm and}\ \  \ A_3 = 0\ee
Comparing these two equations, we see that the instantons indeed
contribute to Berry's phase. The leading order correction to the
asymptotic profile of the monopole is given by the matrix element
\be f = \langle 2| \bar{\psi}_{2-}\psi_{2+}|1\rangle\ee
We now compute the leading contribution to this matrix element.

\subsubsection*{The Instanton Calculation}

Solutions to the kink equations \eqn{bog}  have two collective
coordinates. The first, $T$, simply corresponds to the center of
the kink in Euclidean time. The other collective coordinate is a
phase, arising from the action of the $U(1)_F$ flavour symmetry on
the kink: $\phi_1\rightarrow e^{i\theta}\phi_1$ and
$\phi_2\rightarrow e^{-i\theta}\phi_2$. This phase takes values in
the range $\theta \in [0,\pi)$ since $\phi_i\rightarrow -\phi_i$
coincides with a gauge transformation. In the instanton
calculation we must integrate over these collective coordinates
with a measure obtained by changing variables in the path
integral. Explicitly,
\be \int d\mu_B = \int
\frac{dT}{\sqrt{2\pi}}\sqrt{g_{TT}}\int_0^\pi\frac{d\theta}{\sqrt{2\pi}}
\sqrt{g_{\theta\theta}}\ee
In the appendix we compute the Jacobian factors $g_{TT}=2mr$ and
$g_{\theta\theta} = 2r/m$.

\para
A similar Jacobian factor arises for the two fermionic zero modes.
Both  are Goldstino modes, arising from the action of
supersymmetry on the kink,
\be \lambda_-=-i(\partial_\tau X) \epsilon_-\ \ \ &,&\ \ \
\bar{\psi}_{i+}=\sqrt{2}{\cal D}_\tau \phi_i^\dagger \epsilon_-
\nn\\ \bar{\lambda}_+=i(\partial_\tau X)\bar{\epsilon}_+\ \ \ &,&\
\ \ \psi_{i-}=-\sqrt{2}{\cal D}_\tau \phi_i \bar{\epsilon}_+\ee
The corresponding contribution to the instanton measure is
\be \int d\mu_F = \int d\epsilon_-
\,d\bar{\epsilon}_+\,J_F^{-1}\ee
where the fermionic Jacobian $J_F=2mr$ is computed in the
Appendix.

\para
Finally, in any semi-classical calculation, one must perform the
Gaussian integrals over all non-zero modes around the background
of the instanton. We are used to these cancelling due to
supersymmetry \cite{thooft}. Indeed, the spectra of non-zero
eigenvalues of $\Delta\Delta^\dagger$ and $\Delta^\dagger\Delta$
are equal. However, the spectra are also continuous, and the
densities of eigenvalues need not match. This fact that there is
indeed a mismatch in the densities manifests itself in the index
theorem counting kink zero modes \cite{keith}: one must use the
Callias index theorem, rather than the Atiyah-Singer index
theorem. The net result is that the Gaussian integrals give rise
to a non-trivial contribution in the background of the
kink\footnote{Essentially the same physics is responsible for the
mass renormalization of these kinks in $d=1+1$ dimensional
theories \cite{nick,rebhan}.}. A similar effect was previously
seen in instanton calculations in $d=2+1$ dimensional theories
\cite{dkmtv}.  The explicit computation of the one-loop
determinants is relegated to the Appendix. There we show that
\be {\rm dets} = \sqrt{\frac{\det \Delta
\Delta^\dagger}{\det{}^\prime \Delta^\dagger\Delta}} = 8m
\label{dets}\ee
where the prime denotes the removal of the zero modes.  We are now
almost done. Putting everything together, the instanton
contribution to the matrix element is given by
\be f = \langle 2| \bar{\psi}_{2-}\psi_{2+}|1\rangle  = r \int dT
\int \frac{d\epsilon_-d\bar{\epsilon}_+}{2mr}\,8m \,
|\sqrt{2}{\cal D}_\tau \phi_2|^2 \,e^{-2mr}\ee
Here the integral is to be evaluated on the kink solution. In the
limit $g^2 \rightarrow \infty$, the kink equations \eqn{bog} are
easily solved. In $A_0=0$ gauge, we have
\be \phi_1 =
\frac{\sqrt{r}e^{m\tau}}{\sqrt{e^{2m\tau}+e^{-2m\tau}}}\ \ \ ,\ \
\ \phi_2 = \frac
{\sqrt{r}e^{-m\tau}}{\sqrt{e^{2m\tau}+e^{-2m\tau}}}\ \ \ ,\ \ \ X
= \frac{1}{r} \sum_i m_i |\phi_i|^2\ee
It is now a trivial matter to perform the integral. We find
\be f= 4mr e^{-2mr}\ee
This is merely the leading order correction to the monopole
profile. Indeed, this is not sufficient to explain the smoothness
at the origin \eqn{smooth}. Further corrections presumably arise
from higher-loop effects around the background of the kink. It
would be interesting to understand if the restrictions due to
supersymmetry are strong enough to determine these corrections
exactly\footnote{{\bf Note Added:} In subsequent work, we showed that 
the restrictions due to supersymmetry are indeed strong enough to 
determine the full connection \cite{holonomy}: for the present example, 
the exact Berry connection is given by the BPS 't Hooft-Polyakov monopole 
\cite{bpsberry}.}.

\section*{Appendix: Instanton Calculus}
\setcounter{section}{1} \setcounter{equation}{0}
\renewcommand{\theequation}{\Alph{section}.\arabic{equation}}

In this appendix we collect together the various elements of the
instanton computation of Section 4.

\subsubsection*{Bosonic Jacobian}

The Jacobian for the bosonic collective coordinates is determined
by the overlap of zero modes. Our first task is to compute these
zero modes. The zero modes satisfy the linearized Bogomolnyi
equations,
\be \partial_\tau(\delta X) &=& g^2 \sum_{i=1}^2 (\delta
\phi_i\phi_i^\dagger + \phi_i\delta\phi_i^\dagger)\nn\\  {\cal
D}_\tau \delta \phi_i -i\delta A_0 \phi_i &=& \sum_{i=1}^2
(X-m_i)\delta\phi_i+\delta X \phi_i\ee
We must supplement these with a gauge-fixing condition. We choose
to work in $A_0=0$ gauge, and require,
\be {\cal F}\equiv -\partial_\tau(\delta A_0) + ig^2
\sum_{i=1}^2\left(\phi_i\delta\phi_i^\dagger -
\delta\phi_i\phi_i^\dagger\right) =0\label{gauge}\ee
The bosonic Jacobian $J_B$ is given by the overlap of zero modes,
$J_B=\sqrt{\det g_{\alpha\beta}}$ with
\be g_{\alpha\beta} = \int d\tau \frac{1}{g^2}\left( \delta_\alpha
X \,\delta_\beta X + \delta_\alpha A_0\,\delta_\beta A_0\right) +
\sum_{i=1}^2
\delta_{(\alpha}\phi_i\,\delta_{\beta)}\phi_i^\dagger\ee
We now examine each zero mode in turn.

\subsubsection*{\underline{${\rm Translational\ Mode}$}}

The translational zero modes are given, as usual, by
differentiating the kink solution with respect to $\tau$. If we
work in $A_0=0$ gauge, no further compensating gauge
transformation is necessary. We have
\be \delta\phi_i=\partial_\tau\phi_i \ \ {\rm and}\ \ \ \delta X =
\partial_\tau X\ee
The normalization is given by
\be g_{TT}= \int d\tau\ \frac{1}{g^2}(\partial_\tau X)^2 + 2\sum_i
|{\cal D}_\tau \phi_i|^2 = 2mr \label{jac}\ee
where, to evaluate the integral, we use the instanton equations
\eqn{bog} to express it in terms of the Euclidean action of the
kink.

\subsubsection*{\underline{${\rm Orientational\ Modes}$}}

The orientational modes arise from acting on the solution with the
$U(1)_F$ global symmetry $\phi_1\rightarrow e^{i\theta}\phi_1$ and
$\phi_2\rightarrow e^{-i\theta}\phi_2$. These are to be
compensated by a gauge transformation $\phi_{1,2}\rightarrow
e^{i\beta}\phi_{1,2}$. The infinitesimal transformations are
\be \delta \phi_1 = i\theta\phi_1 + i\beta\phi_1\ \ ,\ \ \
\delta\phi_2 = -i\theta\phi_2 + i\beta\phi_2 \ \ ,\ \ \  \delta
X=0 \ \ , \ \ \   \delta A_0=\partial_\tau\beta\ee
These satisfy the two linearized instanton equations, but the
requirement of the gauge fixing condition \eqn{gauge} gives us an
equation for $\beta$,
\be \partial_\tau^2 \beta = 2g^2(\beta+\theta)|\phi_1|^2 +
2g^2(\beta-\theta)|\phi_2|^2\ee
Noting that this coincides with the second order equation of
motion for $X$, it is once again solved by the kink profile:
$\beta = -\theta X/ m$. The normalization is
\be g_{\theta\theta} = \int d\tau\
\frac{1}{g^2}\frac{(\partial_\tau X)^2}{m^2}+
2(1-X/m)^2|\phi_1|^2+2(1+X/m)^2|\phi_2|^2 =
\frac{2r}{m}\label{jacky}\ee
where, once again, we employ the instanton equations \eqn{bog} to
rewrite the integral as the action.

\subsubsection*{Fermionic Jacobian}

The two fermionic zero modes carried by the kink arise as
Goldstino modes from broken supersymmetry. Explicitly, they are
given by
\be \lambda_-=-i(\partial_\tau X) \epsilon_-\ \ \ &,&\ \ \
\bar{\psi}_{i+}=\sqrt{2}{\cal D}_\tau \phi_i^\dagger \epsilon_-
\nn\\ \bar{\lambda}_+=i(\partial_\tau X)\bar{\epsilon}_+\ \ \ &,&\
\ \ \psi_{i-}=-\sqrt{2}{\cal D}_\tau \phi_i \bar{\epsilon}_+\ee
The fermionic Jacobian $J_F$ is computed by the overlap of these
modes,
\be J_F&=&\int d\tau d\epsilon_- d\bar{\epsilon}_+\
\frac{1}{g^2}\lambda_-\bar{\lambda}_+ + \bar{\psi}_{i+}\psi_{i-}
\nn\\ &=& \int d\tau \ \frac{1}{g^2}(\partial_\tau X)^2 +2 |{\cal
D}_\tau \phi_i|^2 = 2mr\ee

\subsubsection*{Determinants}

%
%

We will compute the one-loop determinants around the background of
the kink. As in the main text, we work with the masses
$\vec{m}\rightarrow (0,0,m)$ and in the gauge $A_0=0$. The
fermionic Dirac operators in the background of the kink are given
in \eqn{dops}. Performing the Gaussian Grassmann integration over
$\lambda$ and $\psi_i$ yields
\be \Gamma_F =
\frac{{\det}{\Delta^\dagger}\,{\det}^{\prime}{\Delta}}{\det
\Delta^\dagger_0\Delta_0}= \left[\frac{
{\det}(\Delta{\Delta^\dagger})\,
{\det}^{\prime}({\Delta^\dagger}\Delta)}{\det^2(\Delta^\dagger_0\Delta^\dagger_0)}
\right]^{1/2}\ee
where the prime denotes removal of the zero modes, and $\Delta_0$
is the vacuum Dirac operator, for which
$\Delta^\dagger_0\Delta_0=\Delta_0\Delta^\dagger_0$. To determine
the bosonic determinants, we expand all fields around the
background profile of the kink;
\be {\phi}_i \rightarrow {\phi}_i +{\delta{\phi}}_i\ \ \ ,\ \ \
A_0 \rightarrow {\delta{A}}_0 \ \ \ ,\ \ \  \vec{X}\rightarrow
({\delta{X}}_1,{\delta{X}}_2, X+{\delta{X}}_3). \ee
The fact that the kink solves the equations of motion ensures that
terms linear in fluctuations vanish in the expansion of the
action. The terms quadratic in fluctuations split into two groups,
with no mixing between them. It can be checked that the quadratic
fluctuation operator for $\delta{\phi}_i$ and $\delta{X}_3$
coincides with the fermionic operator $\Delta^\dagger\Delta$. (The
zero modes of this operator are simply the bosonic zero modes
discussed above). Meanwhile, the remaining two scalars
$\delta{X}_1$ and $\delta{X}_2$ are governed by the fluctuation
operator $-\partial_\tau^2 + 2g^2|\phi_i|^2$. Thus the Gaussian
integration over bosonic fields yields,
\be \Gamma_B = \frac{\det(-\partial_\tau^2 +
2g^2r)\,\det(\Delta^\dagger_0\Delta_0)} {\det(-\partial_\tau^2 +
2g^2|\phi_i|^2)\,\det^\prime(\Delta^\dagger\Delta)}\ee
Finally, we must deal correctly with the gauge symmetry of the
theory using the gauge fixing condition \eqn{gauge}. We implement
this through the standard Fadeev-Popov trick. Under the gauge
transformation $\delta A_0=\partial_\tau{\alpha}$ and
$\delta\phi_i=i\alpha\phi_i$, we have
\be\delta {\cal F} = -\partial_\tau^2\alpha + 2g^2
|\phi_i|^2\alpha\ee
Correspondingly, we introduce ghosts $c$ and $\bar{c}$ with action
\be S_{ghost} = \int \,d\tau\ {\bar{c}}\, (-{\partial}_\tau^2 +
2g^2 |{\phi}_i|^2)\,c\ . \ee
Integrating over the ghosts in the background of the kink, we find
\be \Gamma_{\rm ghost} = \frac{\det(-\partial_\tau^2 +
2g^2|\phi_i|^2)}{\det(-\partial_\tau^2 + 2g^2r)}\ee
which precisely cancels the contribution due to the $\delta X_1$
and $\delta X_2$ scalar fluctuations. The upshot of these
calculation is that the total determinants are given by
\be {\rm dets} = \Gamma_F\Gamma_B\Gamma_{\rm ghost} =
\sqrt{\frac{\det\Delta\Delta^\dagger}{\det^\prime\Delta^\dagger\Delta}}\ee
as advertised in \eqn{dets}.

\para
The operators $\Delta\Delta^\dagger$ and $\Delta^\dagger\Delta$
share the same spectrum of non-zero eigenvalues, but this spectrum
is continuous and the densities of eigenvalues differ. This means
that the ratio of determinants does not cancel. We may calculate
this ratio using the technique of \cite{dkmtv}. We first define
the regulated index for the Dirac operator,
\be {\cal I}(\mu^2) = \Tr\left(\frac{\mu^2}{\Delta^\dagger\Delta +
\mu^2}\right) - \Tr\left(\frac{\mu^2}{\Delta\Delta^\dagger +
\mu^2}\right)\ee
In the limit $\mu^2\rightarrow 0$, ${\cal I}(\mu^2)$ computes the
index of the Dirac operator $\Delta$. However, here we are more
interested in the $\mu$ dependence of the index, due to the
identity,
\be \int_\mu^\infty \frac{1}{\mu^\prime}{\cal I}(\mu^{\prime 2}) =
\frac{1}{2}\log \det \left(\frac{\Delta\Delta^\dagger +
\mu^2}{\Delta^\dagger\Delta + \mu^2}\right)\ee
We strip off the two zero modes to get the primed determinant by
writing
\be \det{}^\prime \Delta^\dagger \Delta = \lim_{\mu^2\rightarrow
0} \frac{1}{\mu^2}\,\det(\Delta^\dagger\Delta + \mu^2)\ee
which allows us to write our desired ratio of determinants as
\be {\rm dets} = \sqrt{\frac{\det
\Delta\Delta^\dagger}{\det{}^\prime\Delta^\dagger\Delta}} =
\lim_{\mu\rightarrow 0} \mu \exp\left(\int_{\mu}^\infty
d\mu^\prime \frac{{\cal
I}(\mu^{\prime2})}{\mu^\prime}\right)\label{ohboy}\ee
Thus it remains only to compute ${\cal I}(\mu^2)$. In fact this
calculation was done some time ago by Lee to compute the number of
zero modes of the most general kinks \cite{keith}. (The
calculation follows closely the index theorem for magnetic
monopoles \cite{erick}). Lee showed that
\be {\cal I}(\mu^2) = \frac{2m}{\sqrt{4m^2 + \mu^2}}\ee
Evaluating this on \eqn{ohboy} gives our desired expression for
the determinants,
\be \sqrt{\frac{\det
\Delta\Delta^\dagger}{\det{}^\prime\Delta^\dagger\Delta}} = 8m \ee
%

%
%
%
%
%

\section*{Acknowledgement}
We would like to thank Nick Dorey, Ken Intriligator, Nick Manton,
Bernd Schroers and Paul Townsend for useful discussions.  D.T.
thanks the Isaac Newton Institute for hospitality during the
completion of this work. C.P. is supported by an EPSRC
studentship. J.S. is supported by the Gates Foundation and STFC.
D.T. is supported by the Royal Society.

\end{document}